\documentclass[twocolumn,aps,prb,reprint,superscriptaddress]{revtex4-2}
\usepackage{amsmath}
\usepackage[utf8]{inputenc}
\usepackage{txfonts}
\usepackage{amssymb}
\usepackage{gensymb}
\usepackage{graphicx}
\usepackage{xcolor}

\usepackage[colorlinks=true, urlcolor=blue, linkcolor=blue, citecolor=blue]{hyperref}
\usepackage[english]{babel}
\makeatletter

\renewcommand{\selectlanguage}[1]{}
\def\eV{{\,\textrm{eV}}}
\def\meV{{\,\textrm{meV}}}
\def\K{{\,\textrm{K}}}

\definecolor{lb}{rgb}{0.000, 0.400, 1.000}

\makeatother

\begin{document}
\title{Large anisotropic magnetoresistance in $\alpha$-MnTe induced by strain}

\author{Bao-Feng Chen}
\affiliation{School of Physical Science and Technology, 
Soochow University, Suzhou 215006, China}
\author{Jie-Xiang Yu}
\email{jxyu@suda.edu.cn}
\affiliation{School of Physical Science and Technology, 
Soochow University, Suzhou 215006, China}

\author{Gen Yin}
\email{gen.yin@georgetown.edu}
\affiliation{Department of Physics, Georgetown University, Washington, D.C. 20057,
USA}

\begin{abstract}
     $\alpha\textrm{-MnTe}$ is a p-type semiconducting altermagnet with a N\'eel temperature near $300\thinspace\textrm{K}$. Due to the altermagnetic nature, $\mathcal{PT}$ symmetry is broken, and Kramers degeneracy is lifted in the valence band maxima along the $\Gamma\textrm{-K}$ line and the A point. However, the energy difference is found to be small, and any small shift in the spectrum can dramatically change the linear-response transport properties. Here we show that a strain modulating the [0001] axis of the unit cell by $\sim\pm0.5\%$ can significantly change the transport signature by switching the thermal window between the two regions of the valence band. When the $\Gamma\textrm{-K}$ line is dominating, the planar Hall effect and the anisotropic magnetoresistance can be enhanced by an order of magnitude, with the maximum up to $\sim70\%$. 
\end{abstract}

\maketitle

%
%
Antiferromagnets attracted extensive attention in the past decade due to their compensated magnetization and ultrafast spin dynamics\cite{jungwirth_antiferromagnetic_2016, baltz_antiferromagnetic_2018, reimers_direct_2024}. 
These features make them ideal candidates for next-generation information-processing platforms such as recording or memory devices. 
However, many antiferromagnets have time-reversal symmetry (combined with spacial symmetry operations), such that Kramers degeneracy persists throughout the Brillouin zone.
This makes it difficult to extract the compensated spin order using carrier transport. 
Therefore, antiferromagnets with lifted Kramers degeneracy become important, represented by many exciting discoveries including the $\textrm{Mn}_3\textrm{X}$ family (non-colinear)\cite{nagamiya_triangular_1982, brown_determination_1990, markou_noncollinear_2018, khadka_kondo_2020, chen_anomalous_2021, gao_epitaxial_2022} and altermagnets (colinear)\cite{lee_theory_2009, ma_multifunctional_2021,smejkal_emerging_2022,cheong_altermagnetism_2025}. 
Among these materials, $\alpha$-MnTe is particularly exciting since it is a semiconducting altermagnet with a N\'{e}el temperature of $\sim300~\K$. 
Previous experiments have demonstrated a sizable magnetoresistance when rotating the N\'eel vector among three equivalent axis \cite{kriegner_multiple-stable_2016,kriegner_magnetic_2017}. 
This has been attributed to the lifted Kramers degeneracy near the valence band maxima (VBM) along the $\Gamma\textrm{-K}$ line induced by spin-orbit coupling\cite{yin_planar_2019, krempasky_altermagnetic_2024,lee_broken_2024}. 
Nevertheless, the energy of the $\Gamma\textrm{-K}$ pockets and the A point are found to be very close, and are highly sensitive to details such as the lattice constant, the spin configuration and the disorder\cite{faria_junior_sensitivity_2023,belashchenko_giant_2025}. 
Recently, compact models have been constructed based on first-principles calculation and symmetry analysis, suggesting that the VBM is dominated by the $p$ orbitals of the Te sites \cite{yin_planar_2019, takahashi_symmetry_2025}. 
Specifically, the $\Gamma\textrm{-K}$ line is dominated by $p_z$ orbitals, whereas the A point is dominated by $p_x$ and $p_y$. 
This suggests that strain may be a dominating factor impacting the overall transport signature. 
Here we show that strain along [0001] can modulate the magnetoresistance by an order of magnitude, resulting in a giant anomalous magnetoresistance (AMR) up to $\sim 70\%$. 
This offers an intriguing experimental handle to manipulate the altermagnetic feature of the spectrum for various spintronic applications including tunneling junctions, Hall effects and spin torques. 
%

%
%
We calculate the linear-response conductivity of $\alpha\textrm{-MnTe}$ using the spectrum near the Fermi energy ($\epsilon_F$). 
Since the Berry-curvature singularities near the valence band top cancel each other, here we only consider the anisotropic transport of charge due to the $\mathbf{k}\textrm{-space}$ geometry of the 3-dimensional (3D) spectrum. 
We start from the nonequilibrium average velocity written as $\langle\mathbf{v}_{g}\rangle=\frac{1}{\rho}\sum_{\mathbf{k},n}f_{1}\mathbf{v}_{g}$, 
where $\rho$ is the carrier concentration, $f_1$ is the linear-response correction of the distribution function and $\mathbf{v}_g=-\frac{1}{\hbar}\nabla_{\mathbf{k}}\epsilon$ is the group velocity. 
The summation is performed over the Brillouin zone for all the bands indexed by $n$. 
Relaxation time approximation is assumed: $f_1=\tau e\frac{\partial f_{0}}{\partial\epsilon}\boldsymbol{{\cal E}}\cdot\mathbf{v}^{k}$, where $f_0(\mathbf{k},n,\epsilon _F)$ is the equilibrium Fermi function and $\boldsymbol{{\cal E}}$ is a uniform external electric field. 
At zero temperature, $\frac{\partial f_0}{\partial \epsilon }\approx-\delta(\epsilon_\mathbf{k}-\epsilon_F)$.
Further considering the linear response of the current: $\langle\mathbf{j}\rangle=\rho e\langle\mathbf{v}_{g}\rangle=\left[\sigma\right]\boldsymbol{{\cal E}}$, we have the Ohmic conductivity tensor as
\begin{equation}
    [\sigma]=-\frac{e^{2}\tau}{\left(2\pi\right)^{3}\hbar^{2}}\left(\sum_{n}\oint_{\epsilon_{n}=\epsilon_{F}}\frac{1}{\left|\nabla_{\mathbf{k}}\epsilon_{n}\right|}\nabla_{\mathbf{k}}\epsilon_{n}\left(\nabla_{\mathbf{k}}\epsilon_{n}\right)^{T}ds\right),\label{eq:conductivityTensor}
\end{equation}
which is a Fermi-surface integration\cite{yu_discrete_2021} capturing both the AMR and PHE. 
The anisotropic electron spectrum used in Eq. \ref{eq:conductivityTensor} is obtained by the localized Wannier function (WF) based Hamiltonian of the converged ground-state electron density given by density functional theory (DFT). 
The DFT calculation is performed using 
projector augmented wave (PAW) pseudopotentials~\cite{blochl_projector_1994, kresse_ultrasoft_1999},
implemented in the Vienna ab initio simulation (VASP) package~\cite{kresse_efficient_1996, kresse_efficiency_1996}.
The generalized gradient approximation in Perdew, Burke, and Ernzerhof (PBE)
formation~\cite{perdew_generalized_1996} was used as the exchange-correlation energy.
An energy cutoff of 600 eV was used for the plane-wave expansion
throughout the calculations.
The $\Gamma$-centered $12\times12\times8$ $k$-mesh are sampled in the Brillouin zone for self-consistent calculations. 
The experimental lattice parameters of $a=4.171$\AA~ and $c=6.686$\AA~\cite{kriegner_multiple-stable_2016} are used.
Spin-orbit coupling was included when magnetic anisotropic spin ordering was considered. 
Based on the eigenstates and eigenvalues obtained from DFT,
a unitary transformation from the plane-wave basis to the WF basis 
was performed to construct the tight-binding Hamiltonian 
by using the band disentanglement method~\cite{souza_maximally_2001}
implemented in the Wannier90 package~\cite{mostofi_updated_2014}.
Mn($3d$) and Te($5p$) orbitals are chosen for the projection. 
For all states below the VBM 
except for core and semi-core electrons such as Te($5s$) electrons,
the WF-based Hamiltonian has exactly the same eigenvalues as those obtained by DFT.

\begin{figure}
\includegraphics[width=1\columnwidth]{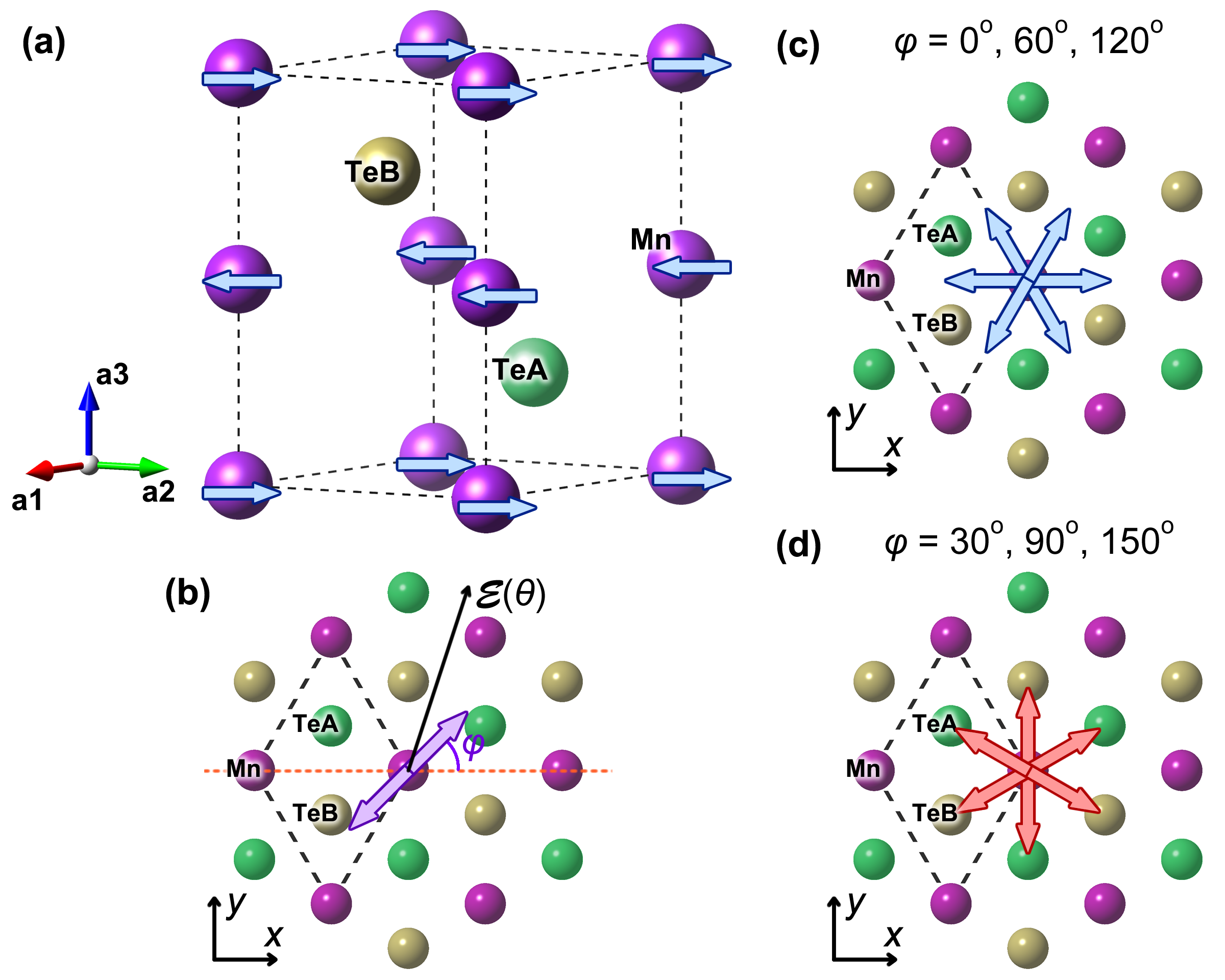}
\caption{The crystal structure of $\alpha$-MnTe illustrated in (a) the main view and (b-d) the top views. 
Purple, Green and yellow balls represent Mn atoms and the two distinct Te sublattice sites, respectively. 
Arrows in (a) display the 3-dimensional colinear antiferromagnetic ordering. 
In (b), $\varphi$ and $\theta$ represents the angle definition of the N\'{e}el vector and the electric fields, respectively. 
(c) and (d) show two groups of the N\'{e}el vectors, each containing three equivalent directions. }
\label{fig:1} 
\end{figure}

\begin{figure}
\includegraphics[width=1\columnwidth]{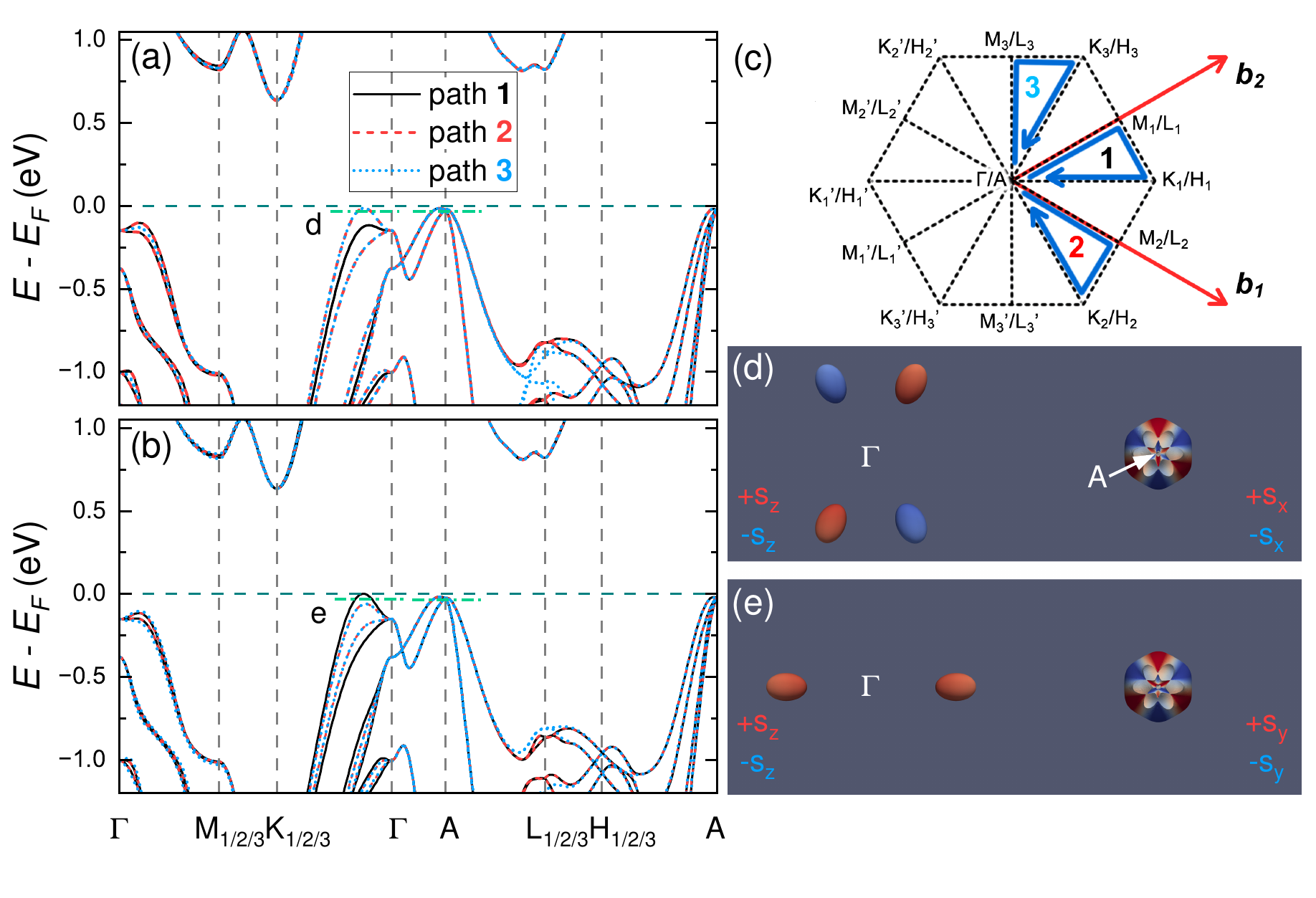}
\caption{The SOC band structure when the N\'{e}el vector is aligned along (a) $\varphi=0\degree$ and (b) $\varphi=90\degree$ with $U=3.0\eV, J=0.9\eV$.
VBM is set to zero.
(c) the Brillouin zone of MnTe with all high symmetric $k$-points: $\Gamma$, $M$ and $K$ are in $k_z=0$ plane and  $A$, $L$, $H$ are in $k_z=\pi/c$ plane.
The $k$-point path for band plotting is also displayed in (b).
(d-e) the Fermi surfaces corresponding to the case of (a) and (b),  respectively. 
The Fermi energy is $0.02\eV$ lower than VBM. 
Red and blue indicate the spin components in the directions labeled on the side.}
\label{fig:bands-fs} 
\end{figure}

%
$\alpha$-MnTe has NiAs-type lattice structure and 
its nonsymmorphic space group is $G=\mathrm{P}6_{3}/\mathrm{mmc}$ (No. 194). 
As shown in Fig.~\ref{fig:1}(a), both Mn and Te form hexagonal lattices in the plane.
Previous studies\cite{kriegner_multiple-stable_2016, kriegner_magnetic_2017} suggested easy plane magnetic anisotropy with the [0001] direction being the hard axis. 
Within the easy plane, the in-plane anisotropy is negligibly small ($< 0.05\meV$ per unit cell). 
In principle, the N\'{e}el vector can thus be aligned along any in-plane direction labeled by the angle $\varphi$ with respect to $+x$, as illustrated in Fig.~\ref{fig:1}(b).
$\varphi\in[0,\pi)$ is therefore enough to describe $\mathcal{T}$-even transport properties here.
Particularly, in both cases of $\varphi=0\degree$ and $\varphi=90\degree$, the N\'eel vector has three equivalent orientations as shown in Figs.\ref{fig:1}(c) and (d), respectively.
The position of VBM is sensitive to the computational functional choice, so that
we tune the VBM position by employing the Hubbard $U$ method in the Liechtenstein implementation~\cite{liechtenstein_density-functional_1995} 
on those Mn($3d$) orbitals to include the on-site correlation of the localized $3d$ electrons.
Figs.\ref{fig:bands-fs}(a) and (b) show the band structures of MnTe with $U=3.0\eV$ and  $J=0.9\eV$ along the $k$-point path labeled in Fig.~\ref{fig:bands-fs}(c). 
The energy extrema of the valence band around the $\Gamma\textrm{-K}$ line and the A point are almost equal.
When $U=4.0\eV, J=0.9\eV$ is used, VBM is located near the $\Gamma\textrm{-K}$ line.
When no Hubbard $U$ is applied, VBM is located near the A point.
Thus, we use these two sets of parameters to investigate the transport properties due to the $\Gamma\textrm{-K}$ line and the A point individually.
Figs.\ref{fig:bands-fs}(d) and (e) show the hole pockets of Fermi surfaces with the N\'{e}el vector aligned along $\varphi=0\degree$ and $\varphi=90\degree$.
Here, we label the two types of Fermi surfaces as the $\Gamma$-top and the $A$-top cases, respectively.
%
%
To capture the Fermi surfaces under hole doping, we set the $\epsilon_F$ at $0.02\eV$ below the VBM in both cases.
In the $\Gamma$-top case, the Fermi surface has different geometry when the N\'{e}el vector is aligned along $\varphi=0\degree$ and $\varphi=90\degree$.
For $\varphi=0\degree$, four hole pockets are symmetrically distributed across the four quadrants, mirroring each other with respect to $k_x=0$ and $k_y=0$.
The pockets in I/III quadrants have a spin polarization opposite to those in II/IV quadrants and all four pockets are spin polarized along $\pm s_z$.
On the other hand when $\varphi=90\degree$, two hole pockets appear symmetrically at $+x$ and $-x$ along $k_{y}=0$.
Both have the same spin polarization along $+s_z$.
Once time is reversed, the spin polarization is correspondingly reversed to $-s_z$ in this case.
These results by DFT are consistent with the $k\cdot p$ model in Refs.~\cite{yin_planar_2019,takahashi_symmetry_2025}

The Fermi surface for the $A$-top case, on the other hand, has hexagram hole pockets around the $A$ point for both $\varphi=0\degree$ and $\varphi=90\degree$, indicating the limited impact from SOC.
The spin components for both $\varphi=0\degree$ and $\varphi=90\degree$ follow the N\'{e}el vector direction, showing the bulk $g$-wave spin splitting at low symmetric area.
The spin properties are therefore dominated by the non-relativistic altermagnetic feature.

\begin{figure*}
\includegraphics[width=0.9\linewidth]{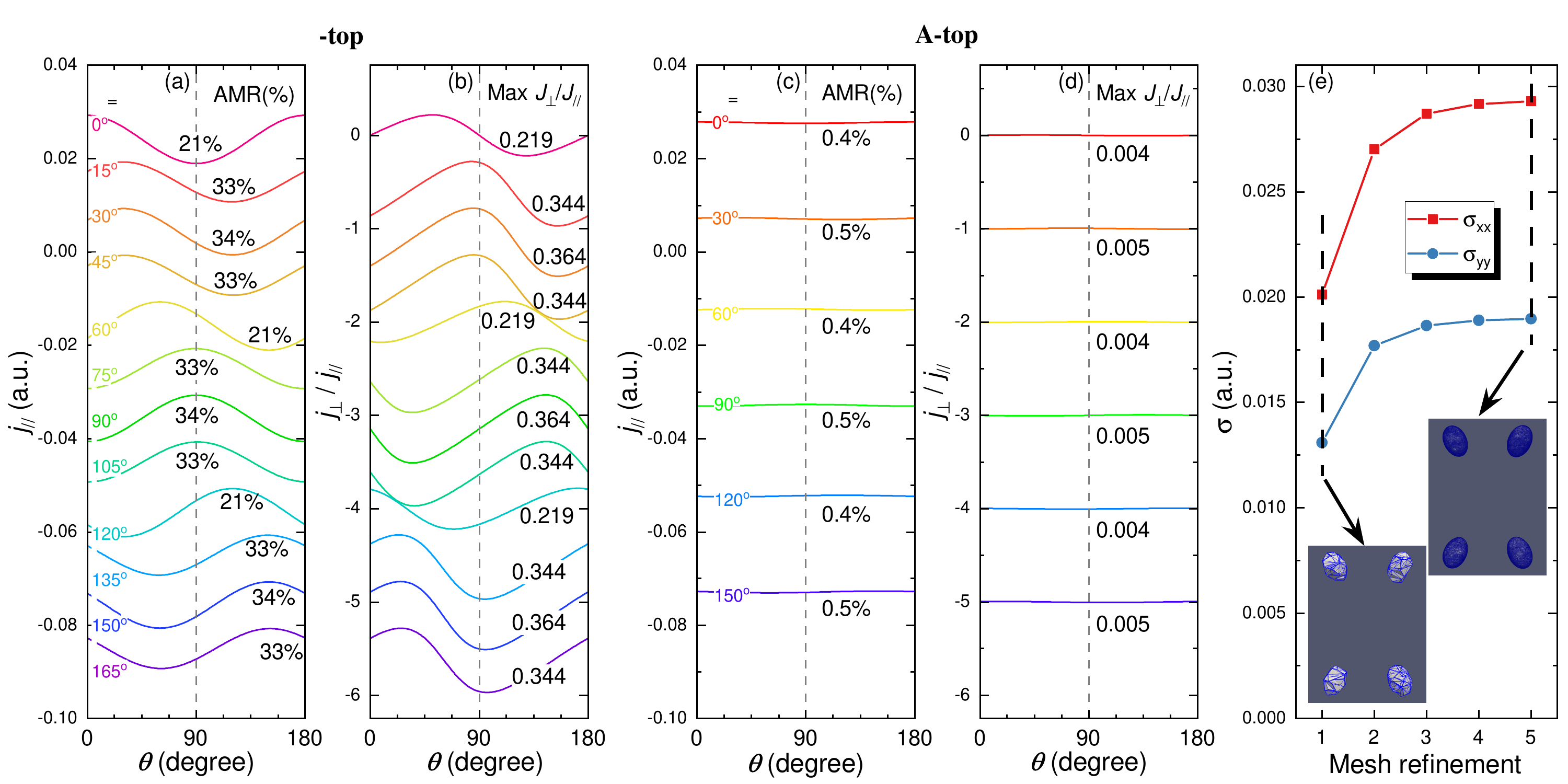}
\caption{The anisotropic magnetoresistance (AMR) and the planar Hall effect (PHE) when rotating the transport direction with respect to the lattice. Here $\theta$ labels the direction of the electric field, whereas $\varphi$ denotes the direction of the N\'eel vector, both with respect to the $+x$ direction as illustrated in Fig. \ref{fig:1}(b). 
Curves in (a) and (b) correspond to the $j_\parallel$ and $j_\perp/j_\parallel$ in the $\Gamma$-top case, respectively. 
Those in (c) and (d) correspond to the $j_\parallel$ and $j_\perp/j_\parallel$ in the $A$-top case, respectively.
The centers of all curves are shifted in the figure
and each $j_\perp/j_\parallel$ curve has the actual central value of zero.
In both cases, the Fermi energy is set to $0.02\eV$ lower than the VBM.
The AMR ratio and the maximum of $j_\perp/j_\parallel$ are marked beside each curve.
(e) the convergence of numerical results of conductivities $\sigma_{xx}$ and $\sigma_{yy}$ after mesh refinement.
The Insets compare the first and the fifth refinement of the fermi surfaces for $\Gamma$-top case with $\varphi=0\degree$.
} 
\label{fig:amr} 
\end{figure*}

%
Due to the negligibly small anisotropy within the easy plane, an in-plane magnetic field can set the N\'{e}el vectors 
along the in-plane direction perpendicular to the magnetic field \cite{kriegner_multiple-stable_2016}.
Since the spectrum geometry is sensitive to the N\'{e}el vector alignment at different angles, the transport properties can be significantly different, resulting in large amplitudes of the AMR and PHE.
We first demonstrate this by calculating the Ohmic conductivity tensor $\sigma$ based on three band structures with N\'{e}el vectors oriented at $\varphi=0\degree$, $90\degree$ and $45\degree$ respectively.
To numerically evaluate Eq. \ref{eq:conductivityTensor}, we first describe the geometry of the Fermi sea by establishing a periodic Delaunay mesh of the Brillouin zone using isotropically random sampling. 
Since the sampling is isotropically random, the mesh does not introduce symmetries that are not intrinsically hosted by the spectrum. 
We then perform mesh refinement by inserting new vertices at the geometric centers in those tetrahedra near the $\epsilon_F$. 
The triangular pieces of the Fermi surface are extracted by interpolating vertices on those edges detected to cross the $\epsilon_F$. 
The geometry of the Fermi surface is therefore approximated by a triangulated curved surface, which is implemented using the exact predicates and constructions implemented by Computational Geometry Algorithms Library (CGAL)\cite{cgal:eb-25b}. 
The triangulated Fermi surface therefore forms a valid manifold with the exact translational symmetry. 
The numerical results are shown in Figs.\ref{fig:amr}(a-e), where Fig.~\ref{fig:amr}(e) displays the values of $\sigma$ in the $\Gamma$-top case for $\varphi=0\degree$ after different rounds of mesh refinements,
confirming the numerical convergence.
%


%
The Ohmic conductivity tensor $\sigma$ is time-reversal symmetric. 
As a result, the conductivity tensors with the N\'{e}el vectors oriented at other equivalent directions $\varphi'$ shown in Figs.\ref{fig:1}(c) and (d) can be obtained by $\sigma(\varphi')=R(\varphi'-\varphi)\sigma(\varphi)R(\varphi'-\varphi)^T$,
where $R$ is the rotation matrix with angle $\varphi'-\varphi$ and $\varphi'-\varphi=\pm\pi/3$.
Similarly, we also have $\sigma(135\degree)=M(90\degree)\sigma(45\degree)M(90\degree)^T$ where $M(90\degree)$ denotes reflection across the $90\degree$ plane.
For each $\sigma(\varphi)$, an in-plane external electric field $\mathcal{E}(\theta)$ is applied along the direction $\theta$ respect to the $+x$ axis shown by the orange dashed line in Fig.~\ref{fig:1}(b).
We can thus obtain the longitudinal and transverse currents relative to the electric field.
The results of the longitudinal currents and the ratio of the transversal to longitudinal currents (Hall angle) at both the $\Gamma$-top and $A$-top are shown in Figs.\ref{fig:amr}(a-d). 
The magnitude of the currents contains the coefficients in Eq. \ref{eq:conductivityTensor} in its unit. 
Although the current depends on the constant relaxation time, the anisotropic transport behavior is solely determined by the geometry of the Fermi surface. 
This is particularly the case when the scattering mechanism is isotropic and local, so that the relaxation time is $\mathbf{k}$-independent. 
In all cases, these results exhibit $180\degree$ periodicity, consistent with symmetry of the Fermi surface.
At the $\Gamma$-top case, the curves of $j_\parallel$ and $j_\perp/j_\parallel$ exhibit significant oscillation 
as a function of the orientation of the electric field. 
The positions and the magnitude of the peaks and dips for $j_\parallel$ and $\frac{j_\perp}{j_\parallel}$ are sensitive to the N\'eel-vector orientation labeled by $\varphi$. 
%
%
The oscillation of the longitudinal current and the Hall angle approaches $34\%$ and $0.36$ at $\varphi=30\degree$, $90\degree$ and $150\degree$, respectively.
On the other hand, at the $A$-top case, both $j_\parallel$ and $j_\perp/j_\parallel$ curves are almost flat with much smaller oscillation.
The anisotropic transport behavior is negligibly small for all N\'{e}el vector orientations.
We therefore conclude that the AMR and PHE in $\alpha-$MnTe are governed by the hole spectrum along the $\Gamma\textrm{-K}$ lines, where the strong spin-orbit coupling is important.
The anisotropic conductivity induced by the Fermi surface geometry is the origin of AMR and PHE.
We consider a very simple two-dimensional system with anisotropic conductivity $\sigma_{yy}=(1-\beta)\sigma_{xx}$ without off-diagonal terms.
With the external electric field $\mathcal{E}$ applied along the direction $\theta$, the current and the corresponding longitudinal and transversal components is given by $\mathbf{j} =  \sigma_{xx}\mathcal{E}[\hat{x}\cos{\theta}+\hat{y}(1-\beta)\sin\theta]$, where $j_\parallel = \sigma_{xx}\mathcal{E}\left(1-\beta\sin^2{\theta}\right)$, and $
j_{\perp} = -\sigma_{xx}~\mathcal{E}~\frac{\beta}{2} \sin{2\theta}$.
As a result, both $j_{\perp}$ and $j_\parallel$ have $180\degree$ periodicity. 
The amplitude of the oscillation corresponding to AMR and PHE is therefore proportional to $\beta$, describing the anisotropy of Ohmic conductivity tensor.
Thus, $\beta$ constitutes the most critical and sole parameter quantifying the magnitude of AMR and the Hall angle in PHE.
We note that the $\pi$-period oscillation is within the uniform-domain approximation. 
Different spin configurations set by field-cooling history or heterostructures in actual samples may support domains with N\'eel vectors along different directions, 
introducing anisotropic scattering and resulting in different oscillation behaviors.

\begin{figure}
\includegraphics[width=1\columnwidth]{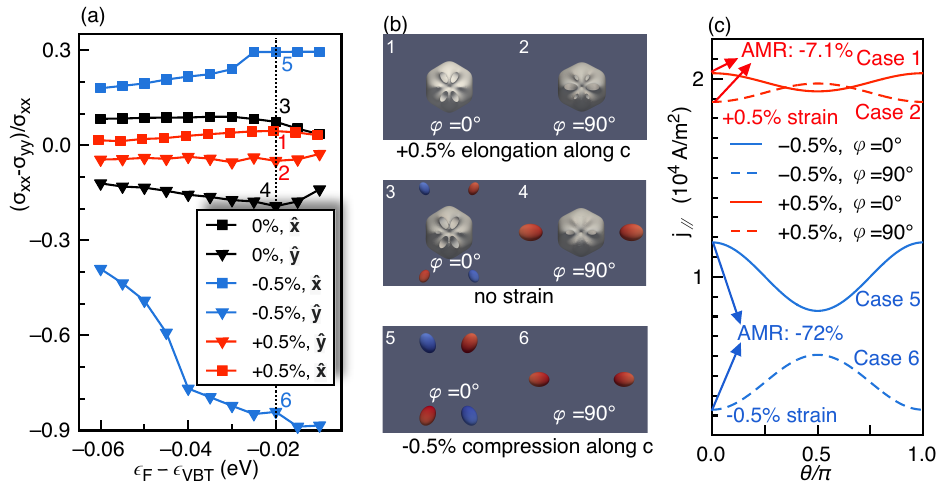}
\caption{(a) Under different magnitudes of the strain along [0001] (the $c$-axis), 
the conductivity anisotropy $\beta=(\sigma_{xx}-\sigma_{yy})/\sigma_{xx}$ as a function of the Fermi energy measured from the VBM.  
Two cases of the N\'{e}el vector $\mathbf{n}=\hat{x}$ ($\varphi=0\degree$) and $\mathbf{n}=\hat{y}$ ($\varphi=90\degree$) are considered, and three different magnitudes of strain ($-0.5\%$, $0\%$ and $+0.5\%$) are demonstrated for each case. 
(b) Fermi surfaces in all the 6 cases with the numbering corresponding to the ones in panel (a). (c) The AMR longitudinal current density $j_\parallel$ as a function of the angle of the electric field. Cases 1 and 2 (elongation) and Cases 5 and 6 (compress) are illustrated. Dashed lines correspond to the case of $\mathbf{n}=\hat x$ and the solid lines are for $\mathbf{n}=\hat y$, respectively. }
\label{fig:strain} 
\end{figure}

%
Since the geometry of the two VBMs are quite different, controlling the relative energy difference between the two is important in experiments. 
Orbital decomposition has uncovered that the spectrum along the $\Gamma\textrm{-K}$ line is dominated by $p_z$ orbitals, whereas the one near the A point is dominated by $p_x$ and $p_y$ \cite{yin_planar_2019, takahashi_symmetry_2025}.
The reason that Hubbard $U$ on Mn($3d$) electrons can control the positions of VBM is 
that the ligand field of Te($5p$) is highly sensitive to the on-site energy of Mn, which consequently modifies the relative energy of the $p_z$ and $p_x/p_y$ orbitals.
This suggests that a strain along [0001] can control the relative energy difference between two VBMs.
In our setting of $U=3.0\eV, J=0.9\eV$, 
the VBM around the $\Gamma\textrm{-K}$ line and the A point are almost equal [bands shown in Fig.~\ref{fig:bands-fs}(a)(b)], which is close to the experimental observation in Ref.~\cite{osumi_observation_2024}. 
We thus further apply
compression and elongation strains along [0001] and then examine the transport properties.
Fig.~\ref{fig:strain}(a) shows the conductivity anisotropy with a $\pm0.5\%$ strain along the $c$-axis (or $\mp0.25\%$ in-plane strain).
The conductivity anisotropy $\beta$ for the unstrained structure [Cases 3, 4 in Fig. \ref{fig:strain}(b)] is much smaller than the $\Gamma$-top case.
At $\epsilon_F-\epsilon_\textrm{VBM}=-0.02\eV$, $\beta$ is about 0.07 for $\varphi=0\degree$ and -0.19 for $\varphi=90\degree$.
The corresponding Fermi surfaces [Cases 3 and 4 in Fig.~\ref{fig:strain}(b)] show the coexistence of the $\Gamma\textrm{-K}$ line and A point pockets.
With $+0.5\%$ compression strain along [0001], the magnitude of $\beta$ is significantly enlarged.
At $\epsilon_F-\epsilon_\textrm{VBM}=-0.02\eV$, we have $\beta\approx0.29$ for $\varphi=0\degree$ and $\beta\approx-0.84$ for $\varphi=90\degree$, which is a modulation of more than one order of magnitude.
The corresponding Fermi surfaces [Cases 5 and 6 in Fig.~\ref{fig:strain}(b)] are the $\Gamma\textrm{-K}$ pockets.
Conversely, with an elongation strain along [0001], $\beta$ approaches a tiny magnitude.
The Fermi surface at $\epsilon_F-\epsilon_\textrm{VBM}=-0.02\eV$ [Cases 1 and 2 in Fig.~\ref{fig:strain}(b)] possesses the A-point pockets only.
This suggests that the spectrum within the transport thermal window drastically changes with the strain along [0001], thereby dramatically changing the amplitude of the AMR.
This change in the AMR ratio can be taken as a transport indicator of the $\Gamma$-top and A-top scenarios in different experimental settings. 
To show this quantitatively, we estimate the constant relaxation time $\tau$ within Drude theory based on the resistivity of $1~\Omega\!\cdot\!\textrm{cm}$ reported in experiments\cite{ferrer-roca_temperature_2000}. 
Assuming $\langle\sigma\rangle = (\sigma_{xx} + \sigma_{yy})/2$, we obtain a
relaxation time of $\tau \approx 2.9\times 10^{-16}~\textrm{s}$. 
Using a representative electric-field magnitude of $100~\textrm{V/m}$, we can
obtain the longitudinal current density as a function of the
electric-field direction $\theta$, which is illustrated in Fig.~\ref{fig:strain}(c). 
This reveals an order-of-magnitude modulation of the AMR ratio $\frac{\rho_\parallel - \rho_\perp}{\rho_\perp}
   = \frac{j_\perp - j_\parallel}{j_\parallel}$
upon applying $\pm 0.5\%$ strain along the [0001] axis. 
Note that the AMR ratio depends on neither the relaxation time nor the magnitude of the
applied electric field within the approximation given by Eq.~\ref{eq:conductivityTensor}. 
Instead, it is determined intrinsically by the geometry
of the hole spectra near the VBM. 
%

%
Our findings suggest that hole spectrum engineering in this altermagnetic semiconductor is essential to implement devices utilizing the lifted Kramers degeneracy.
It should be noted that a moderate in-plane magnetic field -- acting as the driving force for AMR and PHE -- does not directly govern transport properties in $\alpha$-MnTe. 
Its primary role is to reorient the N\'{e}el vector, thereby modulating the Fermi surface geometry. 
As a result, the AMR and PHE originated from the anisotropic Fermi surface is independent from any spin-related scattering,
although the anisotropy itself are influenced by various spin directions via spin-orbit coupling. 
This implies that other factors capable of inducing anisotropy - such as anisotropic external stress or static electric fields - can also generate AMR and PHE, 
which are not inherently spin-dependent.
This reflects a fundamental difference from the $\mathcal{T}$-odd phenomena such as the anomalous Hall effect.
\emph{Acknowledgments:} 
B.-F. C. and J.-X. Y. were financially supported by the National Natural Science Foundation of China (12274309).
G.Y. is supported by the National Science Foundation (US) under Grant \# DMR-2440337. 
The contribution from G.Y. is partially performed on Bridges-2 at Pittsburgh Supercomputing Center through allocation PHY230018 from the Advanced Cyber infrastructure Coordination Ecosystem Services \& Support (ACCESS) program, which is supported by National Science Foundation (US) grants \#2138259, \#2138286, \#2138307, \#2137603, and \#2138296.
\bibliography{main}	

\end{document}